\documentstyle[prl,aps,epsf,here]{revtex}
\twocolumn

\begin{document}

\def\epsi{\varepsilon_i}
\def\d{\hbox{d}}
\newcommand{\Figurebb}[9]{
  \begin{figure}[H]\begin{center}
  \leavevmode
  \epsfysize=#7cm
  \epsfbox[#2 #3 #4 #5]{#6}
  \par
  \parbox{#8cm}{
  \caption[figure]{\renewcommand{\baselinestretch}{0.8} \small
                   \hspace{-0.3truecm}#9}\label{#1}}
  \end{center}
  \end{figure}
}

\title{PERIODIC-ORBIT BIFURCATIONS\\ AND SUPERDEFORMED SHELL STRUCTURE}
\author{A. G. Magner$^{1,2}$, S. N. Fedotkin$^{2}$,
K. Arita$^3$, K. Matsuyanagi$^4$, M. Brack$^5$}

\address{$^1$Research Center for Nuclear Physics, Osaka University,
         Osaka 567-0047, Japan}
\address{$^2$Institute for Nuclear Research, 2523068 Prospekt Nauki 47,
         Kiev-28, Ukraine}
\address{$^3$Department of Physics, Nagoya Institute of Technology,
         Nagoya 466-8555, Japan}
\address{$^4$Department of Physics, Graduate School of Science, Kyoto
         University, Kyoto 606-8502, Japan}
\address{$^5$Institute for Theoretical Physics, University of Regensburg,
         D-93040 Regensburg, Germany}

\maketitle

\vspace*{-.98cm}

\begin{abstract}

We have derived a semiclassical trace formula for the level density of
the three-dimensional spheroidal cavity. To overcome the divergences
occurring at bifurcations and in the spherical limit, the trace integrals
over the action-angle variables were performed using an improved stationary
phase method. The resulting semiclassical level density oscillations and
shell-correction energies are in good agreement with quantum-mechanical
results. We find that the bifurcations of some dominant short periodic
orbits lead to an enhancement of the shell structure for ``superdeformed''
shapes related to those known from atomic nuclei.


\end{abstract}


\vspace*{0.1cm}


{\it Introduction ---} The periodic orbit theory (POT)
\cite{gutzpr,bablo,strumag,bt76,strusem,mfimbrk,book} is a nice tool for
studying the correspondence between classical and quantum mechanics and,
in particular, the interplay of deterministic chaos and quantum-mechanical
behavior. But also for systems with integrable or mixed classical dynamics,
the POT leads to a deeper understanding of the origin of shell structure
in finite fermion systems from such different areas as nuclear 
\cite{strusem,frisk,ak95,fispot}, metallic cluster \cite{nish,erice}, or 
mesoscopic semiconductor physics \cite{qdot,chan}. Bifurcations of periodic 
orbits may have significant effects, e.g., in connection with the so-called 
``superdeformations'' of atomic nuclei \cite{strusem,mfimbrk,ak95,ask98}, 
and were recently shown to affect the quantum oscillations observed in the 
magneto-conductance of a mesoscopic device \cite{chan}.

In the semiclassical trace formulae that connect the quantum-mechanical
density of states with a sum over the periodic orbits of the classical
system \cite{gutzpr,bablo,strumag}, diver\-gences arise at critical
points where bifurcations of periodic orbits occur or where symmetry
breaking (or restoring) transitions take place. At these points the
stationary-phase approximation, used in the semiclassical evaluation of
the trace integrals, breaks down. Various ways of avoiding these divergences
have been studied \cite{bablo,bt76,creagh}, some of them employing uniform
approximations \cite{tgu,sie97,ssun,hhun}. Here we employ an improved
stationary-phase method (ISPM) for the evaluation of the trace integrals
in the phase-space representation, based on the studies in Refs.\
\cite{bt76,sie97}, which we have derived for the elliptic billiard
\cite{mfammsb}. It yields a semiclassical level density that is regular at
all bifurcation points of the short diameter orbit (and its repetitions)
and in the circular (disk) limit. Away from the critical points, our result
reduces to the extended Gutzwiller trace formula
\cite{strumag,strusem,mfimbrk,book} and is identical to that of Berry and
Tabor \cite{bt76} for the leading-order families of periodic orbits.

The main purpose of the present note is to report on the extension of our
semiclassical approach to the three-dimensional (3D) spheroidal cavity
\cite{mfamb}, which may be taken as a simple model for a large deformed
nucleus \cite{strusem,frisk} or a (highly idealized) deformed metal cluster
\cite{nish,erice}, and to specify the role of orbit bifurcations in the
shell structure responsible for the superdeformation. Although the
spheroidal cavity is integrable (see, e.g., Ref.\ \cite{nish91}), it 
exhibits all the difficulties mentioned above (i.e., bifurcations and
symmetry breaking) and therefore gives rise to an exemplary case study of
a non-trivial 3D system. We apply the ISPM for the bifurcating orbits and
succeed in reproducing the superdeformed shell structure by the POT, hereby
observing a considerable enhancement of the shell-structure amplitude near
the bifurcation points.

{\it Theory ---} The level density $g(E)$ is obtained from the semiclassical
Green function \cite{gutzpr} by taking the imaginary part of its trace in
$(\bf I,\Theta)$ action-angle variables \cite{mfimbrk,mfammsb}:
\begin{eqnarray}
g(E) =\sum_i \delta\left(E-\epsi\right)
     \simeq {\rm Re} \sum_\alpha \int {{\rm d}{\bf I}'\,{\rm d}{\bf \Theta}''
     \over {\left(2 \pi \hbar\right)^3}}\, \delta\left(E - H\right)
\nonumber\\
\times\exp\left\{{i \over \hbar}
\left[S_\alpha\left({\bf I}',{\bf I}'',t_\alpha\right)+
\left({\bf I''}-{\bf I}'\right) \cdot {\bf \Theta}''\right] -
i{\pi \over 2} \mu_\alpha\right\}.
\label{gofe}
\end{eqnarray}
Here $\left\{\epsi\right\}$ is the single-particle energy spectrum and
$H=H({\bf I})$ is the classical Hamiltonian. The sum is taken over all
classical trajectories $\alpha$ specified by the initial actions ${\bf I}'$
and final angles ${\bf \Theta}''$. $S_\alpha\left({\bf I'},{\bf I''},
t_\alpha\right)=-\int_{\bf I'}^{\bf I''}{\bf I}\cdot{\rm d}{\bf \Theta}$
is the action integral and $t_\alpha$ the time for the motion along the
trajectory $\alpha$, and $\mu_\alpha$ is the Maslov index related to the
caustic and the turning points \cite{mfammsb,mfamb}. In the spheroidal 
variables $\{u,v,\varphi\}$, the action $\bf I$ has the components
\begin{eqnarray}
I_u &=& {p \,c \over \pi}\int\limits_{-u_c}^{u_c}
        {\rm d}u \, \sqrt{\sigma_1-\sin^2{u}
        -\sigma_2/\cos^2{u}}\,,\nonumber\\
I_v &=& {p \,c \over \pi} \int\limits_{v_c}^{v_t}
        {\rm d}v \, \sqrt{\cosh^2{v}-\sigma_1
        -\sigma_2/\sinh^2{u}}\,,\nonumber\\
I_\varphi &=& p \,c\sqrt{\sigma_2}\,.
\label{iuv}
\end{eqnarray}
Hereby $p=(2mE)^{1/2}$ is the particle's momentum and $c=(b^2-a^2)^{1/2}$
half the distance between the foci; $b$ and $a$ are the semiaxes (with 
$b>a$) of the spheroid with its volume fixed by $a^2b=R^3$ and the axis 
ratio $\eta=b/a$ as deformation parameter; and $\pm u_c$ (or $v_c$) and 
$v_t$ are the caustic and turning points, respectively. In Eq.\ (\ref{iuv}) 
we use the dimensionless ``action'' variables $\sigma_1$, $\sigma_2$  
\cite{mfammsb} in which the torus of the classical motion is given by
\begin{eqnarray}
\sigma_2^- = 0 \,\leq & \,\sigma_2\; & \leq {1\over{\eta^2-1}}=\sigma_2^+,
                   \nonumber\\
\sigma_1^- = \sigma_2 \leq & \,\sigma_1\; & \leq {\eta^2\over{\eta^2-1}}-
                   \sigma_2 \left(\eta^2-1\right) = \sigma_1^+.
\label{bound}
\end{eqnarray}

In the ISPM, we expand the action $S_\alpha$ in (\ref{gofe}) up to 
second-order terms around its stationary points and keep the
pre-exponential factor at zero order, taking the integrations over the
torus within the {\it finite} limits given by Eq.\ (\ref{bound}). For the
oscillating (``shell-correction'') part of the level density $\delta g(E)=
g(E)-{\widetilde g}(E)$, where ${\widetilde g}(E)$ is its smooth part 
\cite{book,strut}, we obtain
\begin{equation}
\delta g(E) \simeq {1\over E_0}\,{\rm Re}\sum_\beta A_\beta(E)
        \exp\left(ikL_\beta-i{\pi \over 2}\mu_\beta\right)w_\beta^\gamma,
\label{tracetotal}
\end{equation}
where $k=p/\hbar$ is the wave number and $E_0=\hbar^2/2mR^2$ our energy
unit. The amplitudes $A_\beta$ will be specified below. The sum over
$\beta$ includes all 2-parameter families of three-dimensional (3D)
periodic orbits and elliptic and hyperbolic 2D orbits lying in a plane
containing the symmetry axis (all with degeneracy parameter $\cal K$=2),
the 1-parameter families of (2D) equatorial orbits lying in the central
plane perpendicular to the symmetry axis (with $\cal K$=1), and the (1D)
isolated long diameter (with $\cal K$=0). In Eq.\ (\ref{tracetotal}),
$L_\beta$ is the length of the orbit $\beta$ at the stationary point
($\sigma_1^*,\sigma_2^*$) which for the 3D orbits lies inside the physical
region of the torus (\ref{bound}), and is analytically continued outside
this region. The $\sigma_2=0$ boundary of (\ref{bound}) is occupied by the
2D orbits with $\cal K$=2. The stationary points are determined by the
roots of the periodicity conditions $\omega_u/\omega_v=n_u/n_v$ and
$\omega_u/\omega_\varphi=n_u/n_\varphi$; hereby $\omega_\kappa=\partial 
H/\partial I_\kappa$ are the frequencies and $n_\kappa$ are co-prime 
integers which specify the periodic orbits $\beta=M(n_v,n_\varphi,n_u)$, 
where $M$ is the repetition number. The factor $w_\beta^\gamma = 
\exp(-\gamma^2L_\beta^2/4R^2)$ in Eq.\ (\ref{tracetotal}) is the result of 
a convolution of the level density with a Gaussian function over a range 
$\gamma$ in the variable $kR$. This ensures the convergence of the POT sum 
(\ref{tracetotal}) by suppressing the longer orbits which are not relevant 
for the coarse-grained gross-shell structure \cite{mfimbrk,book}.

For Strutinsky's shell-correction energy $\delta U$ 
\cite{strumag,book,strut}, we obtain 
(with time reversal symmetry and a spin factor 2)
\begin{eqnarray}
\delta U & = & \; 2 \sum_{i=1}^{N/2} \epsi 
                  - 2 \!\int_0^{{\widetilde E}_F}\! E\, 
                  {\widetilde g}(E)\,{\rm d}E \nonumber\\
    & \simeq & \; 8 R^2 E_F\,{\rm Re}\sum_\beta 
                  {A_\beta(E_F)\over{L_\beta^2}} 
                  \exp\!\left(\!ik_FL_\beta\!-\!i{\pi\over 2}
                  \mu_\beta\!\right).
\label{esc}
\end{eqnarray}
The Fermi energies $E_F$ (and with it $k_F$) and ${\widetilde E}_F$ are 
determined by the particle number conservation $N=2\int_0^{E_F}\!g(E)\,
{\rm d}E=2\int_0^{{\widetilde E}_F}\!{\widetilde g}(E)\,{\rm d}E$. Due to 
the factor $L_\beta^{-2}$, the PO sum in (\ref{esc}) may converge faster for 
the shortest orbits than the level density (\ref{tracetotal}) for small 
$\gamma$. Any enhancement of the amplitudes $A_\beta$ of the most degenerate 
short periodic orbits -- e.g., due to bifurcations or to symmetry restoring, 
as discussed below -- therefore leads to an enhancement of the shell 
structure and hence to an increased stability of the system.

We present here only the amplitudes of the leading contributions to
(\ref{tracetotal}) and (\ref{esc}). For further details (including, e.g.,
explicit expressions for the Maslov indices $\mu_\alpha$), we refer to a
forthcoming, more extensive publication \cite{mfamb}.

For the amplitudes $A_\beta$ of the most degenerate (${\cal K}$=2) families 
of periodic 3D and 2D orbits, we obtain
\begin{equation}
A_\beta^{({\cal K}=2)} = {ic L_\beta\;
                 [\partial I_u/\partial\sigma_1]_{\sigma_n^*}
                 \over {\pi (4M R n_v)^2 \sqrt{K_{\beta}\,\sigma_2^*}}}
                 \,\prod_{n=1}^2
                 \mbox{erf}\left(x_{n}^-,x_{n}^+\right).
\label{amp3d2d}
\end{equation}
The quantity $K_{\beta}=K_\beta^{(1)} K_\beta^{(2)}$ is related to the
main curvatures $K_\beta^{(n)}$ of the energy surface
$E=H(\sigma_1,\sigma_2)$ in the ``action'' plane $(\sigma_1,\sigma_2)$,
given by
\begin{equation}
K_\beta^{(n)} = \left[ {\partial^2 I_v \over {\partial \sigma_n^2}}
                 +{\omega_u \over {\omega_v}}
                {\partial^2 I_u \over {\partial \sigma_n^2}}
                 +{\omega_\varphi \over {\omega_v}}
                {\partial^2 I_\varphi\over {\partial \sigma_n^2}}
                \right]_{\sigma_n^*}\!\!\!\!.
                \quad (n=1,2)
\label{curv}
\end{equation}
In Eq.\ (\ref{amp3d2d}), the arguments of the two-dimensional error 
function erf$\left(x,y\right)= 2 \int_{x}^{y}dz~e^{-z^2}/\sqrt{\pi}$ are 
given by the turning points in the action plane
\begin{equation}
x_n^\pm = \sqrt{-i \pi M n_v K_\beta^{(n)}/\hbar}\,
          \left(\sigma_n^\pm-\sigma_n^*\right);
          \quad(n=1,2)
\label{zvaripm}
\end{equation}
see Eq.\ (\ref{bound}) for the boundaries $\sigma_n^\pm$. All quantities 
in (\ref{amp3d2d}) can be expressed analytically in terms of elliptic 
integrals. For the 3D orbits, our result (\ref{amp3d2d}) is in agreement 
with that obtained by exact Poisson summation over the EBK spectrum (cf.\ 
Refs.\ \cite{bt76,book}).

For the contribution of the ${\cal K}=1$ families of equatorial orbits to
(\ref{tracetotal}), we obtain the amplitudes
\begin{equation}
A_\beta^{({\cal K}=1)}=
\,\sqrt{{i\sin^3\phi_\beta \over {\pi M n_v kR \eta F_\beta}}}\,
\prod_{n=1}^{3}\mbox{erf}\left(x_{n}^-,x_{n}^+\right),
\label{ampeq}
\end{equation}
where $\phi_\beta=\pi n_\varphi/n_v$, $F_\beta$ is 
their stability factor \cite{gutzpr,bablo,mfimbrk}, $\sigma_1^* = 
\sigma_2^*=\cos^2{\phi_\beta}/(\eta^2-1)$, and
\begin{equation}
x_{3}^+ = kc\,\sqrt{{-i \pi\hbar F_\beta \over {64Mn_v (\sigma_2^*+1)
          K_\beta^{(1)}}}}\,,\qquad x_{3}^-=0\,.
\label{zperppmeq}
\end{equation}

The contribution of the isolated long diameter orbit, which may be 
expressed in terms of incomplete Airy integrals \cite{mfammsb,mfamb}, is 
not important for deformations of the order $\eta \sim 1.2 - 2$.

\Figurebb{Fig1}{124}{254}{568}{530}{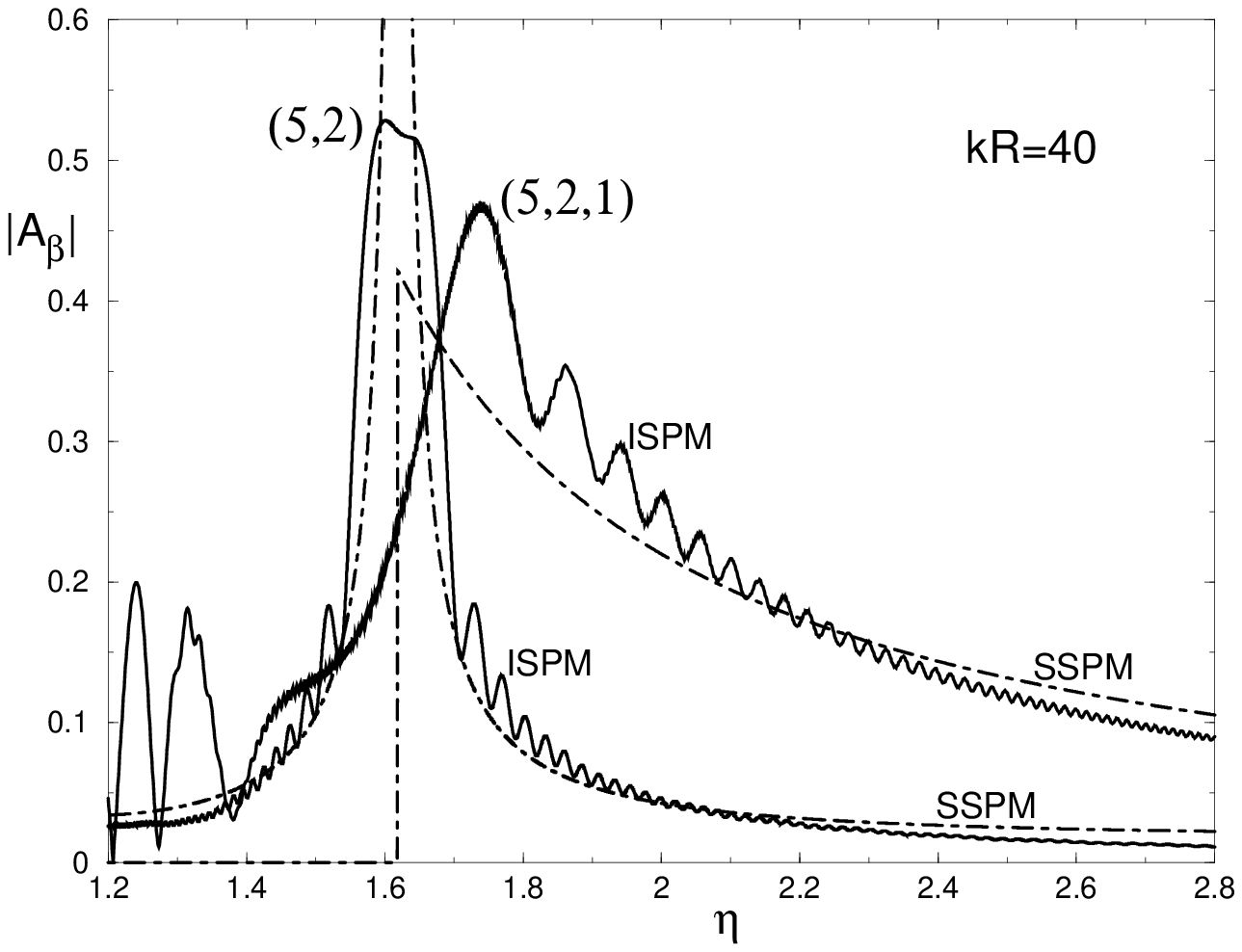}{6.5}{8.5}{
Moduli of amplitudes $|A_\beta|$ versus $\eta$ for the equatorial
``star'' orbit (5,2) (${\cal K}=1$) and the 3D orbit (5,2,1) (${\cal K}=2$) 
bifurcating from it at $\eta$=1.618... {\it Solid lines:} using the ISPM 
according to Eqs.\ (\ref{ampeq}) and (\ref{amp3d2d}), respectively; 
{\it dash-dotted lines:} using the standard stationary-phase approach.
}

\vspace*{-0.25cm}

{\it Discussion of results ---}
In Fig.\ 1 we show $|A_\beta|$ versus deformation $\eta$ (at $kR=40$) for
a pair of orbits involved in a typical bifurcation scenario. At the 
critical point $\eta=1.618...$ the equatorial ``star'' orbit (5,2) 
undergoes a bifurcation at which the 3D orbit (5,2,1) is born; the latter 
does not exist below $\eta=1.618...$

In the standard stationary-phase approach (SSPM; dash-dotted lines), the
amplitude of the (5,2) orbit diverges at $\eta=1.618...$, whereas that of
the bifurcated orbit (5,2,1) is finite but discontinous. As seen in Fig.\ 
1, the ISPM (solid lines) leads to a finite amplitude $A_{(5,2)}^{({\cal 
K}=1)}$ for the (5,2) orbit. This is because the factor $F_\beta$ in the 
denominator of (\ref{ampeq}), which goes to zero at the bifurcation, is 
cancelled by the same factor in the numerator of $x_3^+$ (\ref{zperppmeq}) 
via the third error function in (\ref{ampeq}). A similar result was found 
for the short diameter orbit 2(2,1) in the elliptic billiard \cite{mfammsb}. 
Furthermore, the ISPM softens the discontinuity for the (5,2,1) orbit, 
leading to a maximum amplitude slightly above the critical deformation.

The relative enhancement of these amplitudes $A_\beta$ near the bifurcation
point can also be understood qualitatively from the following argument.
At the bifurcation of the equatorial (5,2) orbit, its degeneracy parameter
${\cal K}=1$ locally increases to 2, because it is there degenerate with 
the orbit family (5,2,1) that has ${\cal K}=2$ at all deformations $\eta 
\geq 1.618...$ This is similar to a symmetry restoring transition. An 
increase of the symmetry parameter ${\cal K}$ by one unit leads to one 
more exact integration compared to the SSPM, and thus the amplitudes 
(\ref{amp3d2d}) and (\ref{ampeq}) acquire an enhancement factor 
$\sqrt{kL_\beta} \propto \sqrt{pR/\hbar}$ (cf.\ Refs.\ \cite{strumag,book}).

A similar enhancement of the double triangle 2(3,1) and the 3D orbit
(6,2,1) is found near their bifurcation point $\eta=\sqrt{3}=1.732...$
However, the curvature $K_\beta^{(1)}$ (\ref{curv}) for orbits like
$M(3t,t,1)$ ($t=2,3$,...) is identically zero and hence the SSPM is
divergent for all deformations $\eta \geq 1$, in contrast to the situation
with orbits like (5,2,1) with finite $K_{\beta}^{(1)}$. Here we have to
take into account the next nonzero 3rd-order terms in the expansion of
$S_\alpha$, although the $(3t,t,1)$ ISPM amplitude (\ref{amp3d2d}) is
finite and continuous everywhere. The amplitude can then be expressed in 
terms of incomplete Airy and Gairy integrals with finite limits 
\cite{mfamb}. For the equatorial orbits $t$(3,1), like for the double 
triangles 2(3,1), one has a zero curvature $K_\beta^{(1)}$ only at 
the bifurcation point $\eta=\sqrt{3}$. Here $F_\beta/K_\beta^{(1)} 
\rightarrow 0$, and a similar mechanism of cancellation of singularities 
for other orbits takes place through Eqs.\ (\ref{zvaripm}-\ref{zperppmeq}). 
But the relative enhancement of the ISPM amplitudes 
(\ref{amp3d2d},\ref{ampeq}) of such orbits at the bifurcations is of order 
$kL_\beta$ because of a change of the degeneracy parameter ${\cal K}$ by 
{\it two} units (see \cite{mfamb} for details). In this sense we avoid 
here a double singularity related to a double restoring of symmetry.

In Figs.\ 2 and 3, we present semiclassical level densities $\delta g(E)$
(\ref{tracetotal}) versus $kR$ and shell-correction energies $\delta U$
(\ref{esc}) versus $N^{1/3}$ for various critical deformations (heavy 
dotted lines), and compare them to the corresponding quantum-mechanical 
results (thin solid lines). We observe a very good agreement of the 
gross-shell structure at all deformations. The most significant 
contributions to these results near the critical deformations are coming 
from bifurcating orbits with lengths smaller than about 10$R$, in line 
with the convergence arguments for the POT sums (\ref{tracetotal}) and 
(\ref{esc}) mentioned above. For the bifurcation at $\eta=1.618...$, the 
orbits (5,2,1) and (5,2) give contributions comparable with other 2D
orbits. For $\eta=\sqrt{3}$, the bifurcating orbits (6,2,1) and (6,2)
are also important.

\Figurebb{Fig2}{132}{245}{568}{574}{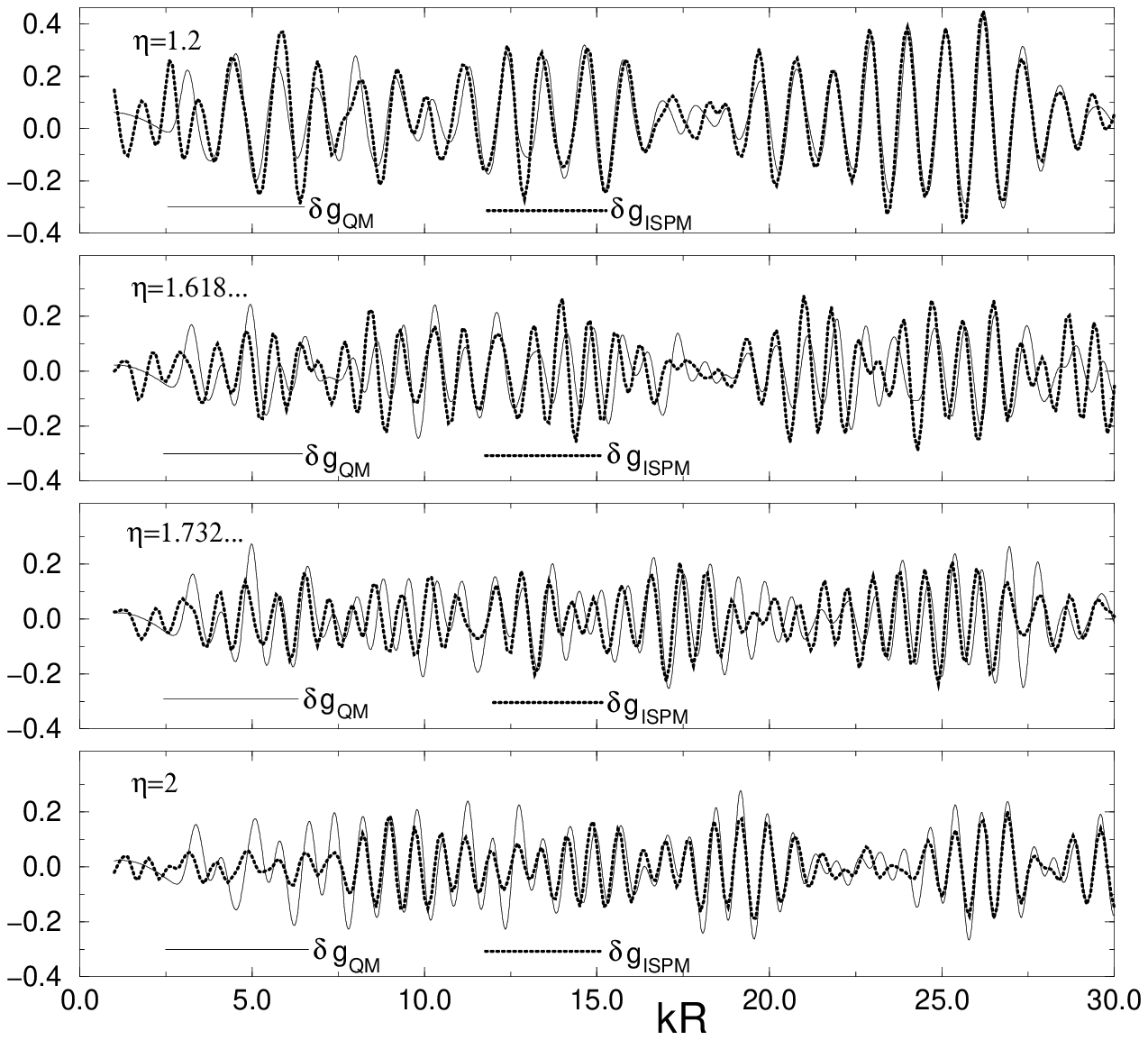}{7.85}{8.5}{
Level density $\delta g(E)$  (\ref{tracetotal}) (unit $E_0^{-1}$) versus 
$kR$ for different critical deformations $\eta$. The Gaussian averaging 
parameter is $\gamma=0.3$. {\it Thin solid lines:} quantum-mechanical 
results; {\it heavy dotted lines:} semiclassical results using the ISPM.
}

\Figurebb{Fig3}{121}{241}{568}{575}{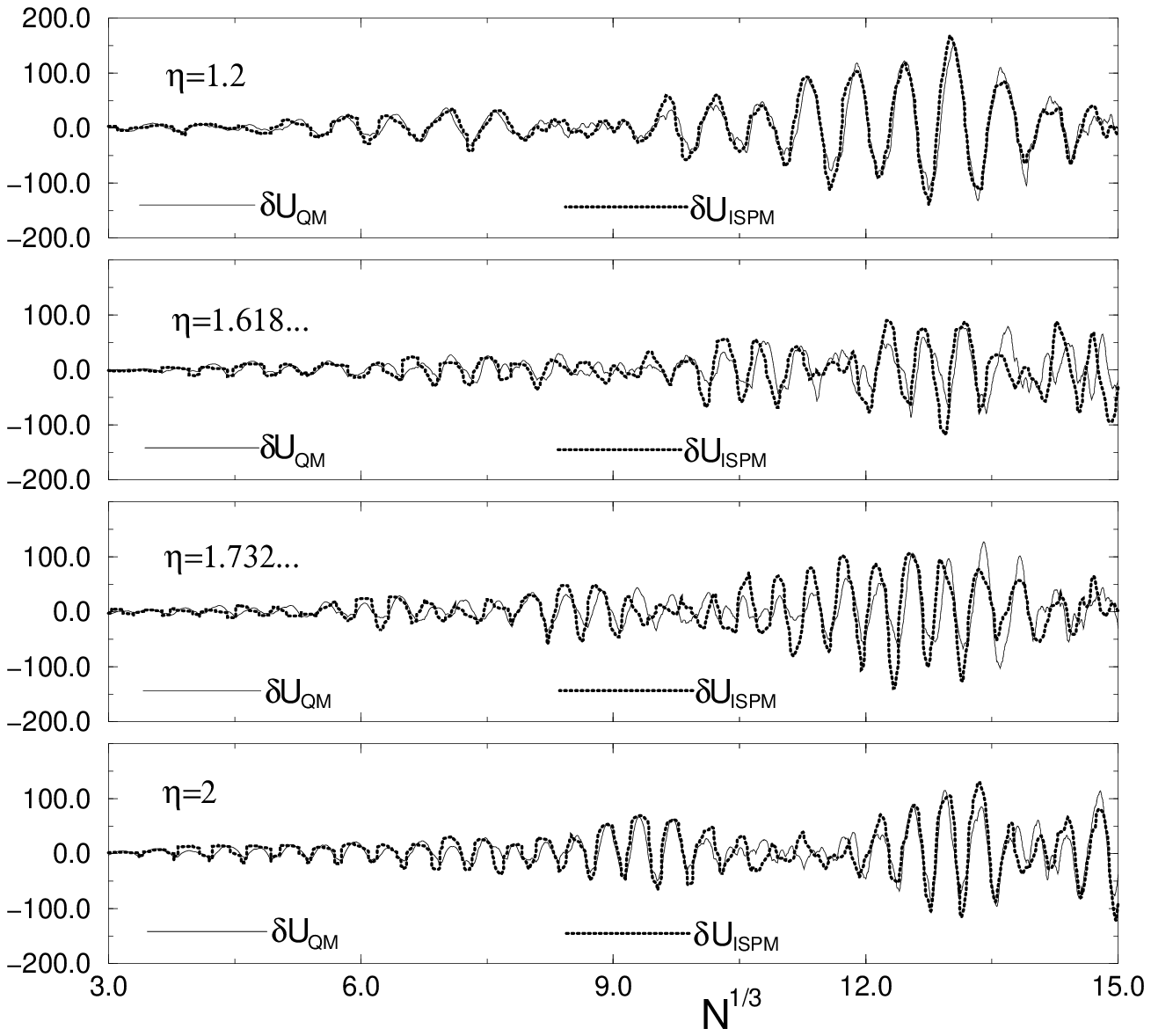}{7.7}{8.5}{
Shell-correction energy $\delta U$ (\ref{esc}) (unit $E_0$) versus cube
root of particle number $N^{1/3}$ (same notation and same deformations
as in Fig.\ 2).
}

\vspace*{-0.3cm}

The role of the bifurcating orbits increases for larger deformations and 
is dominating at the superdeformation $\eta=2$. For this deformation, 
the most important orbits in the present spheroidal cavity model are the 
3D orbits (5,2,1), (6,2,1), (7,2,1), and (8,2,1).

These results are in agreement with both heights and positions of the
peaks in the length spectra obtained in Ref.\ \cite{ask98} from the
Fourier transforms of the quantum level densities $g(kR)$ at the same
deformations.

\vspace*{0.1cm}

{\it Summary and conclusions ---}
We have obtained an analytical trace formula for the 3D spheroidal cavity 
model, which is continuous through all critical deformations where
bifurcations of periodic orbits occur. We find an enhancement of the 
amplitudes $|A_\beta|$ at deformations $\eta\sim 1.6-2.0$ due to 
bifurcations of 3D orbits from the shortest 2D orbits. We believe that
this is an important mechanism which contributes to the stability of
superdeformed systems. Our semiclassical analysis may therefore lead 
to a deeper understanding of shell structure effects in superdeformed 
fermionic systems -- not only in nuclei or metal clusters but also, e.g., 
in deformed semiconductor quantum dots whose conductance and magnetic 
susceptibilities are significantly modified by shell effects.

\vspace*{0.2cm}

{\it Acknowledgements ---} A.G.M.\ gratefully acknowledges the financial 
support provided under the COE Professorship Program by the Ministry of
Education, Science, Sports and Culture of Japan (Monbu-sho), giving him 
the opportunity to work at the RCNP, and thanks Prof.\ H. Toki for his 
warm hospitality and fruitful discussions. Two of us (A.G.M. and S.N.F.) 
acknowledge financial support by the Regensburger Universit\"atsstiftung 
Hans Vielberth.


\end{document}